\documentclass[twocolumn,twoside,floatfix,showpacs,pre,aps]{revtex4}
\usepackage{graphicx,amsmath,amssymb,bm}
\usepackage[colorlinks=false]{hyperref}

\begin{document}

\title{Local dimension and finite time prediction in spatiotemporal chaotic
systems}

\author{Gerson Francisco}
\email{gerson@ift.unesp.br}

\author{Paulsamy Muruganandam}

\affiliation{Instituto de F\'{\i}sica Te\'orica, Universidade
Estadual Paulista, 01405-900 S\~ao Paulo - SP, Brasil}

\date{}

\begin{abstract}

We show how a recently introduced statistics [Patil {\it et al},
Phys. Rev. Lett. {\bf 81} 5878 (2001)] provides a direct
relationship between dimension and predictability in
spatiotemporal chaotic systems. Regions of low dimension are
identified as having high predictability and vice-versa.  This
conclusion is reached by using methods from dynamical systems
theory and Bayesian modelling. We emphasize in this work the
consequences for short time forecasting and examine the relevance
for factor analysis. Although we concentrate on coupled map
lattices and coupled nonlinear  oscillators for convenience, any
other spatially distributed  system could be used instead, such as
turbulent fluid flows.

\end{abstract}


\maketitle

\section{Introduction}

Systems formed by aggregates of parts that interact in a nonlinear
way are prototypes of complex behaviour in physics, biology and
economics. An often successful approach in the analysis of  such
systems is to use tools from dynamical systems theory applied
locally at points of spatially distributed configurations. In many
cases local nonlinearities lead to unpredictable chaotic evolution
where short time forecasts  are still feasible. An instance of
this situation is atmospheric research where, for obvious reasons,
prediction is often the most important goal to be achieved. In
general, for large spatially distributed systems, it is highly
desirable to have a simple diagnostic tool to identify  regions of
predictable behaviour. The main result of this paper is to show
that certain spatial regions do indeed yield better forecasts than
other locations. To elaborate on this theme we use the
concept of BV dimension, introduced in \cite{patil:prl:2001} in
the context of the Earth's atmosphere, as the tool to identify
regions of high dimensionality. Our objective is then to show that
this dimensional estimate is directly related to our ability to
make short time forecasts. A recently introduced Bayesian
approach, the cluster weighted modelling, CWM
\cite{gershenfeld:99-01}, is one of the methods on which we
base our conclusions. We start  from a simple prediction algorithm
\cite{kantz:book:97}  where the main idea of the paper is
readily recognized. The more  sophisticated CWM approach in
conjunction with the simpler prediction algorithm will
provide a clear link between dimension and our ability to
forecast the behaviour of deterministic systems.
In particular, Bayesian approaches also permit the calculation of
the variance (confidence intervals) of forecasts. In Ref.
\cite{salvino:95-01} we find a similar approach where a measure of
predictability is based on the variance under the evolution of
suitably defined functions in embedding space. Such approach
requires long evolution times which cannot be afforded in the
present context and, as will be seen herein, this problem is
circumvented by the use of the CWM.

The concept of correlation dimension \cite{grassberger:83-01,
grassberger:83-02} has some bearing on this subject since high
dimensional systems cannot be easily distinguished from systems
with dominant stochastic component, where predictability  is known
to be low. Here by prediction we mean the existence of a
deterministic map which can be successfully used to make short
time forecasts, in the least square sense or using some more
general likelihood function.

The paper is organized as follows. In Sec. \ref{sec2}, we present
the basic aspects and definitions about the BV dimension, factor
analysis and   the necessary tools to relate dimension and
predictability.  In Sec. \ref{sec3}, the spatiotemporal systems
given by coupled logistic maps and coupled Lorenz systems are
discussed  and the results of the simulations are presented. Our
concluding  remarks are outlined in Sec. \ref{sec4}.

\section{Local dimension and predictability}\label{sec2}

The notion of dimension used herein is based on the concept of
bred vectors. They are constructed in a similar way as Lyapunov
vectors but in practical applications they differ in two aspects.
Firstly, for bred vectors there is no global orthonormalization and
secondly, they are finite amplitude, finite time vectors. Such
properties facilitate the calculation of bred vectors and yield an
efficient identification of regions were short time forecasts are
feasible.

\begin{figure}[!ht]
\begin{center}
\includegraphics[width=0.75\linewidth]{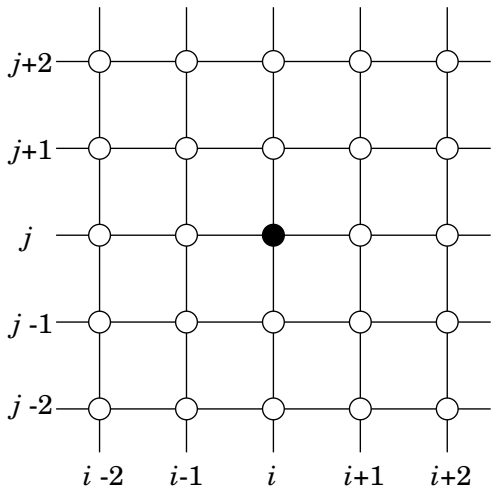}
\end{center}
\caption{Schematic diagram showing the choice of nearest
neighbours at site $(i, j)$ for local dimension. The bred
vectors are the dynamical variables associated with
these sites. } \label{fig_low}
\end{figure}

Consider a 2D spatially distributed system whose state at a given
time $t_1$ is defined over a collection of points $(i,j)$. Here we
take the $M-1$ nearest neighbours for each point $(i,j)$ in a
square lattice with $M=25$, as illustrated in  Fig. \ref{fig_low}.
Logistic maps are one variable dynamical systems and in order to
specify the
corresponding state at a point including its neighbours we need an
$M$ dimensional state vector. Considering the $x-y$ coordinates of
the Lorenz system specified at a suitable constant $z$ hyperplane,
the state vector in this case requires 2$M$ components. In
general the state vector, either $M$ dimensional or 2$M$
dimensional, will be called bred vector. Now generate $k$ distinct
perturbations of the state starting at $t_0 < t_1$ obtaining $k$
local bred vectors. The $k\times k$ covariance matrix of the
system is just $\bm{C} = \bm{B^TB}$, where $\bm{B}$ is the
$M\times k$ (or $2M\times k$ for the Lorenz system) matrix of
local bred vectors each normalized to unity. In this paper we will
fix $k=5$.

We order the eigenvalues of the covariance matrix as  $\lambda_1
\ge \lambda_2 \ge \ldots \ge \lambda_k$ and define the  singular
values of $\bm B$ as $\sigma_i = \sqrt{\lambda_i}$. Here the
connection with factor analysis is clear \cite{johnson:book:98},
since the eigenvalues of the covariance matrix give an idea of the
local linear independence of the $k$ local bred vectors. An
effective dimension of the space of bred vectors can be obtained
by fixing a threshold value corresponding to the highest $l$
eigenvalues, as is done in principal components analysis. Thus an
approximation of the data, supposing zero average for simplicity,
is contained in the product $\bm{\mathcal{F}L}$ were
$\bm{\mathcal{F}} = \bm{BL}^T$ is called the factor and  $\bm{L}$
the loading matrix of dimension $l\times k$. The rows of the
loading matrix are the components of the eigenvectors
corresponding to the dominant eigenvalues. Clearly there is an
arbitrariness in the stipulation of $l$ and this ambiguity is
absent when using the concept of BV dimension.

The eigenvalues $\lambda_i$ represent the amount of variance in
the set of the $k$ unit bred vectors. In order to estimate
unambiguously the value of the threshold one defines the following
statistic \cite{patil:prl:2001}
\begin{equation}\label{eq:bvd}
\psi_{i,j}(\sigma_1, \sigma_2,\ldots,\sigma_k) =
\frac{\left(\sum_{l=1}^{k}\sigma_l\right)^2}{\sum_{l=1}^{k}\sigma_l^2}
\end{equation}
As each of the $k$ bred vectors is normalized to unity, $\psi$
assumes values in the interval $(0,k)$. Examples of the values of
this statistic for several distributions of  bred vectors can be
found in \cite{patil:prl:2001}. A property of the statistic just
defined is its robustness under noise or numerical errors. It can
be used to determine the dominant eigenvalues $l$: just take it to
be the smallest integer bigger than $\psi$. In this sense an
approximation to the bred vectors is obtained as the product of
the corresponding factor and the loading matrix.

We describe now the criteria and tools that will be implemented in
the next section to spatially distributed systems in order to
obtain a connection between dimension and predictability. We start
with a simple prediction algorithm which is very useful in the
analysis of nonlinear time series \cite{kantz:book:97}. To
describe this criterion, consider a scalar time series obtained
from a component of the dynamical variables, say, the
$x$-component, defined at a given site in the 2D lattice. Using
delay coordinates one considers neighbours to within a distance
smaller than $\delta$ to a given point in embedding space. For the
present analysis the choice of embedding dimension 3 is adequate.
By propagating the corresponding neighbours of a given vector in
embedding space and averaging them we obtain a forecast of the
vector. The more sophisticated CWM \cite{gershenfeld:99-01} is a
Bayesian  approach where the embedded time series is used to build
local Gaussian models. One of the advantages of this method is
that it goes beyond point prediction since it includes error bars
(confidence intervals) when estimating the future average value.
The idea is to expand the joint probability distribution
$p(\mathsf{y}, \vec {\mathsf x})$ in terms of local Gaussian
models. Using the $x$ components at each site for definiteness,
the value to be predicted $\mathsf{y}$ is taken as $x(n+2\tau)$
(e.g. with embedding dimension $m=3$ and delay $\tau$), given the
vector of delayed components $\Vec{\mathsf{x}} =
\{x(n),x(n+\tau),n\}$ . The joint distribution is used to compute
the average predicted values, $\langle {\mathsf y}|\vec {\mathsf
x} \rangle$ and their variances, $\langle {(\mathsf y}-\langle
{\mathsf y}|\vec {\mathsf x} \rangle)^2|\vec {\mathsf x} \rangle$.
This provides a criterion for predictability since higher variance
is associated to lower predictability. The use of uncertainty as
measured by the variance is employed in \cite{salvino:95-01} to
obtain a definition of predictability associated to chaotic
systems or stochastic processes and such method is used to
distinguish between these two modes of evolution. Such distinction
has also been made in the context of time series analysis using
the CWM approach \cite{ferreira:02-01}.

\section{Local dimension in Spatiotemporal chaotic systems}\label{sec3}

\subsection{Coupled logistic maps}

We begin with the case of two dimensional coupled map lattices
consisting of logistic maps \cite{bohr:prl:1989}.
\begin{eqnarray}
\label{eqn:cml}
x^{i,j}_{n+1} & = & (1-\epsilon)f\left(x^{i,j}_{n}\right)
+\frac{\epsilon}{4}
\left[ f\left(x^{i-1,j}_{n}\right) \right.\nonumber \\
& & \left.
+f\left(x^{i+1,j}_{n}\right)
+f\left(x^{i,j-1}_{n}\right)
+f\left(x^{i,j+1}_{n}\right)\right],
\end{eqnarray}
with
\begin{equation}\label{log:map}
 f(x) = \mu x(1-x), \;\;\;\;x\in(0,1),\;\;\;\;\mu\in(0,4),
\end{equation}
where $i,j = 1,2,\ldots N$ and $\epsilon$ represents the coupling
strength.

In this simulation we use $N=50$, $\mu=4$, $\epsilon=0.4$ and
periodic boundary conditions. The reference spatial variables
$x^{i,j}$, $i,j=1,2\ldots N$, are obtained by evolving system
(\ref{eqn:cml}) from random initial conditions. The number of
iterations is chosen to be 5015 so that transients are removed.
\begin{figure}[!ht]
\begin{center}
\includegraphics[width=0.8\linewidth]{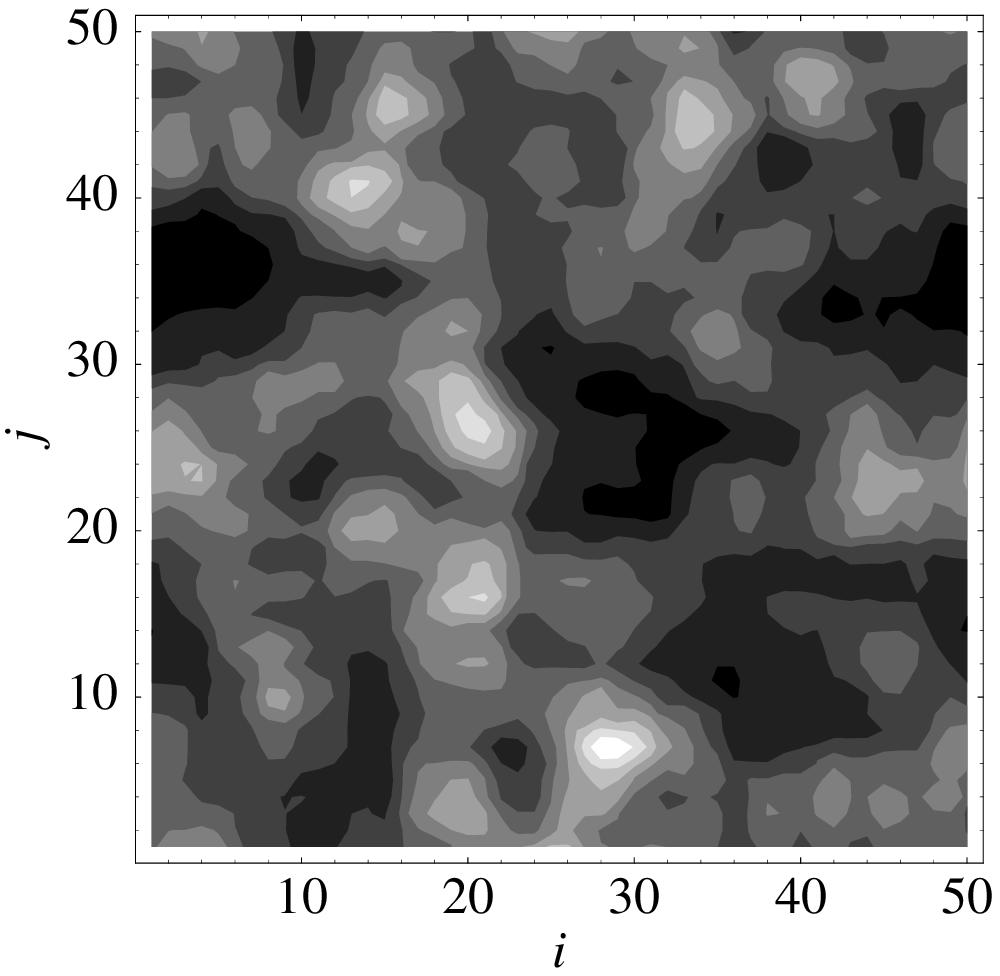}
\end{center}
\caption{Gray scale plot showing the regions of low (dark) and
high (bright) local dimensions for the coupled logistic maps
(\ref{eqn:cml}).} \label{cml_bvd}
\begin{center}
\includegraphics[width=0.9\linewidth]{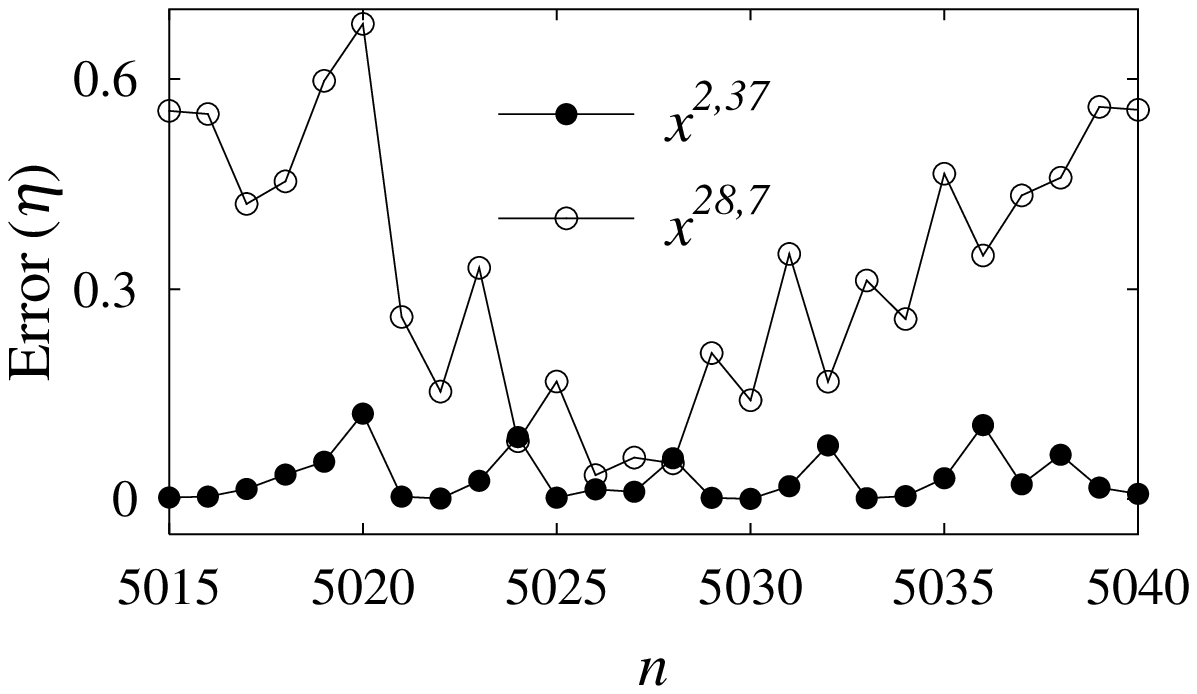}
\end{center}
\caption{Simple finite prediction error for the coupled logistic
maps. The error remains minimum value for $x^{2,37}$ (low BVD
point) and high values of error obtained for $x^{28,7}$ (high BVD
point).} \label{pre_log}
\end{figure}We generate additional spatial variables by adding small
perturbations to the reference variables at time 5000 to measure
the local instability of the coupled maps. Thus at time  5015 we
have spatial distributions corresponding to the reference variable
and four distinct perturbations. The local dimension at the
spatial points  are computed at this time value using the
statistic  defined in Eq. (\ref{eq:bvd}) as discussed in the
previous section. Fig.~\ref{cml_bvd} illustrates the  results of
the logistic maps (\ref{eqn:cml}) calculations where dark regions
correspond to low dimensions and bright  regions represent high
dimensions. We found that  the local dimension has a minimum
value, $\psi=1.0323$, at $i=2$, $j=37$ and a maximum value,
$\psi=2.3636$, at $i=28$, $j=7$. We note that the maximum and
minimum values at the spatial points $(i,j)$ are practically
constant under evolution of system (\ref{eqn:cml}) a few steps
forwards or backwards.

In order to establish a connection between dimension and
predictability, we analyse the time series at the points $(2,37)$
and $(28,7)$ discussed above. We apply the simple prediction
algorithm during the interval from $n=5015$ to $n=5040$. Fig.
\ref{pre_log} shows the prediction errors, $\eta=\mid
x^{i,j}_{\mbox{orig}} - x^{i,j}_{\mbox{pred}}\mid$, for $x^{2,37}$
and  $x^{28,7}$ with $\delta = 0.05$. It is evident that the
prediction error is small at the lattice point $(2,37)$ where the
local dimension is low, and large at the point $(28,7)$ where the
local dimension is maximum. In addition, similar results were
obtained for points with other local dimension values. We found
that the prediction error is consistently  low in regions of low
local dimension and high in the regions of  relatively high local
dimension.

The Bayesian approach known as the cluster weighted modelling will
provide more insights into the predictability issue. Particularly
important is the relationship between variance of predicted
\begin{figure}[!ht]
\begin{center}
\includegraphics[width=0.9\linewidth]{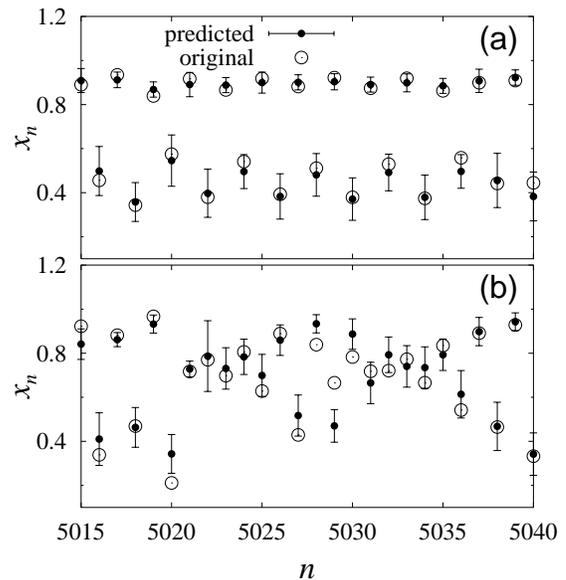}
\end{center}
\caption{Prediction of time series for the coupled logistic maps
by using cluster weighted modelling. The original (circles) and
predicted values (filled circles with errorbars) of the  time
series at (a) low BVD location $(2,37)$ and (b) high BVD location
$(28,7)$.} \label{cwm_log}
\end{figure}
average value and the BV dimension. We consider the average
prediction $\langle {\mathsf y}|\vec {\mathsf x} \rangle$ using
the conditional distribution obtained from the joint distribution
$p({\mathsf y},\vec{\mathsf x})$. As discussed in the previous
section, we take ${\mathsf y}$ to be the $x$ variable at time
$n+2\tau$ and $\vec {\mathsf x} = \{x_n,x_{n+\tau},n\}$; here we
take $\tau=1$. Fig. \ref{cwm_log} presents a series of predicted
values, using always the two most recent original values, and the
corresponding variance. In most cases analysed we found the
following behaviour. Lattice sites with high BV dimension result
in predicted values with larger variances than those with low BV
dimension, or either the prediction tend to fall outside the
confidence interval defined by the variance of the future value.
These facts are shown in Figs. \ref{cwm_log}(a) and (b) for the
low and high BV dimensions, respectively 1.0323 and 2.3636. In
this case the simple prediction results are more compelling than
the CWM since the difference between highest and lowest BV
dimension is not big enough. In the next subsection we discuss
another system where the maximum BV dimension is about four times
larger than the minimum dimension over the spatial distribution.
In this case  the conclusion that the uncertainty in prediction is
related to dimension is even more forceful.

\subsection{Coupled Lorenz systems}

We next consider a two dimensional array of diffusively coupled
Lorenz oscillators represented by the following equations
\cite{kacarev:prl:96}:
\begin{eqnarray}
\dot x_{i,j} & = & \sigma (y_{i,j}-x_{i,j})+c(y_{i-1,j}+
y_{i+1,j}+y_{i,j-1}
\nonumber \\ &   & + y_{i,j+1}-4y_{i,j}), \nonumber \\
\dot y_{i,j} & = & - x_{i,j}z_{i,j}+Rx_{i,j}-y_{i,j}, \nonumber \\
\dot z_{i,j} & = &   x_{i,j}y_{i,j}-\beta z_{i,j}, \label{eqn:lor}
\end{eqnarray}
where $i,j=1,2,\ldots, N$, $\sigma=16$, $R=40$, $\beta=4$ and
$c=1$.  We use $50\times 50$ ($N=50$) oscillators with periodic
boundary conditions.  Equations (\ref{eqn:lor}) are solved
numerically using fourth order Runge-Kutta method with random
initial conditions. We actually consider the $x$ and $y$
components representing a map of the above system by the
Poincar\'e section taken at $z=R-1$. The spatial distribution of
the local dimensions is calculated in a similar fashion as in the
case of coupled map lattices discussed earlier. A low value of
$\psi = 1.1447$ is found at lattice point $(i,j)=(3,50)$, while a
\begin{figure}[!ht]
\begin{center}
\includegraphics[width=0.8\linewidth]{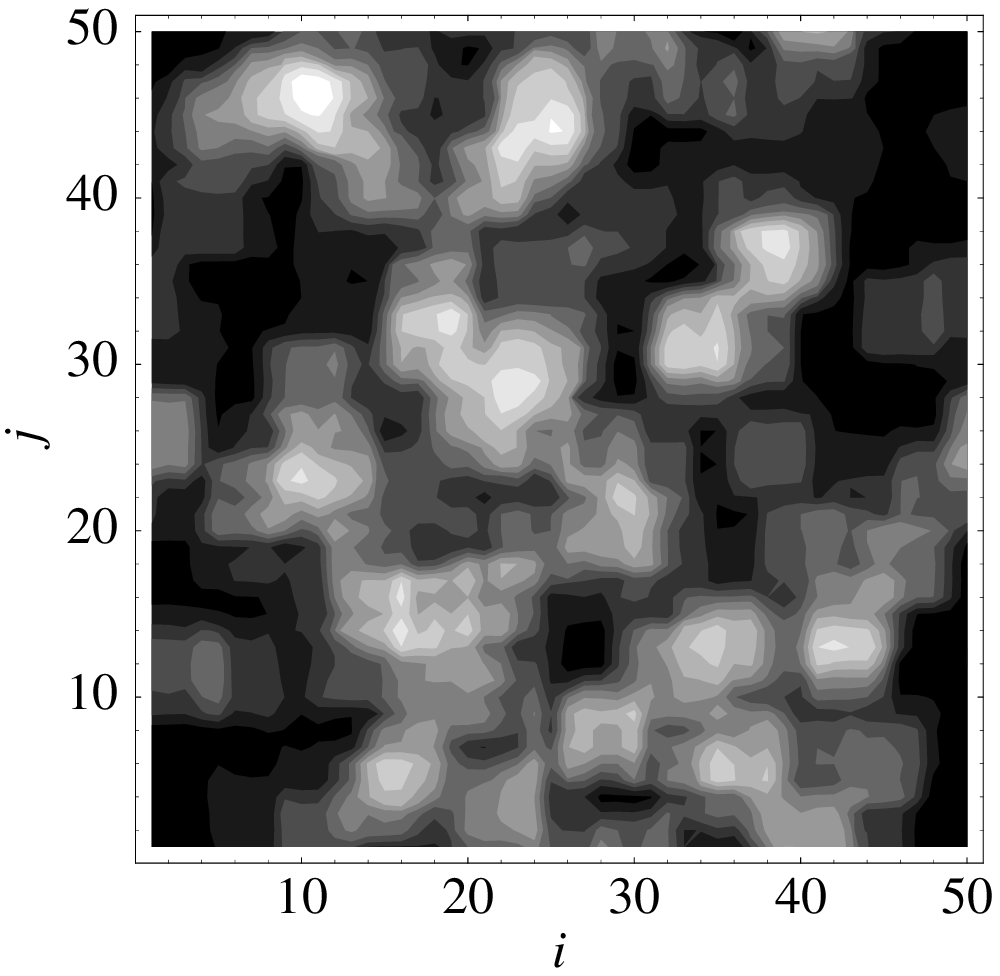}
\end{center}
\caption{Gray scale plot showing the regions of low (dark) and
high  (bright) local dimensions for the coupled Lorenz systems
(\ref{eqn:lor}).} \label{lor_bvd}
\begin{center}
\includegraphics[width=0.9\linewidth]{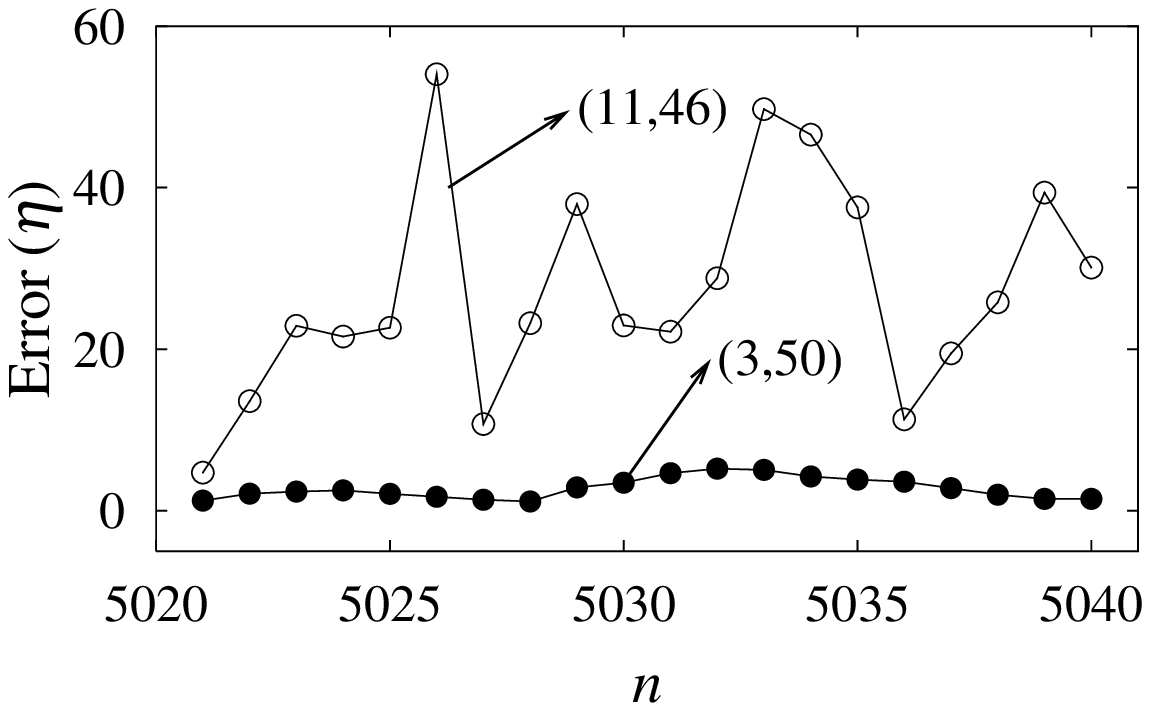}
\end{center}
\caption{Simple finite time prediction error for the coupled
Lorenz  systems. The absolute error remains minimum at $(3,50)$
for which the BV dimension is low and high values of error
obtained at $(11,46)$, the high BV dimension point.}
\label{lor_pre}
\end{figure}
\begin{figure}[!ht]
\begin{center}
\includegraphics[width=0.9\linewidth]{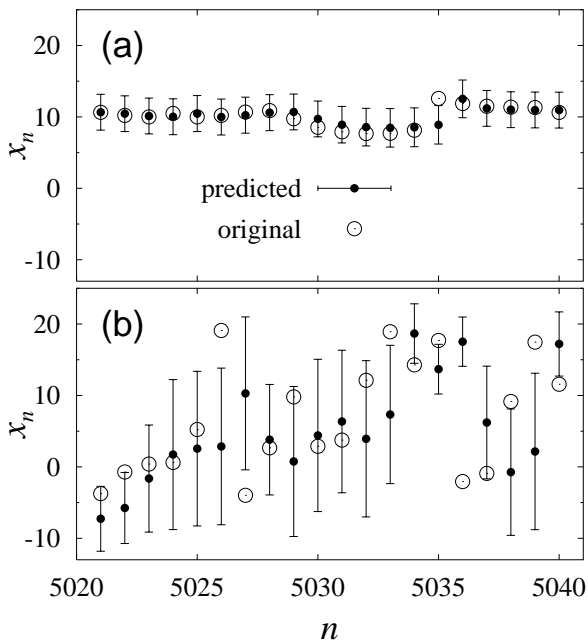}
\end{center}
\caption{Prediction of the time series using cluster weighted
modelling at (a) $(3,50)$ and (b) $(11,46)$.} \label{lor_cwm}
\end{figure}
high value $\psi = 4.1226$ is computed at $(i,j)=(11,46)$. Fig.
\ref{lor_bvd} shows the regions of different local dimension for
the coupled Lorenz systems (\ref{eqn:lor}). The finite time
prediction error in Fig. \ref{lor_pre} uses $\delta\approx 0.8$ in
the simple prediction algorithm. Here the error is calculated as
$\eta = \sqrt{ \left( x_{\mbox{orig}} - x_{\mbox{pred}} \right)^2
+ \left( y_{\mbox{orig}} - y_{\mbox{pred}} \right)^2}$. The
results concerning the cluster weighted modelling are shown in
Fig. \ref{lor_cwm}. Here high local dimension imply a consistent
increase in the variances represented by larger error bars. In
this case there is a clear relationship between predictability and
the local BV dimension.

\section{Summary and conclusions}\label{sec4}

Predictability is related to the uncertainty in the outcome of
future events during the evolution of the state of a system. We
have interpreted the CWM as a tool to detect such uncertainty and
used it in spatially distributed systems. The simple prediction
algorithm in conjunction the CWM form a powerful set of methods to
relate predictability and dimension. Another tool based on the
variance of future states of a system is also employed in
\cite{salvino:95-01} where a level of predictability is defined
and applied to distinguish between deterministic and stochastic
behaviour. Such distinction requires a propagation time longer
than the short time behaviour used herein and the cluster weighted
modelling is more appropriate in the present context. Both
methods, however, CWM or \cite{salvino:95-01}, are more than
diagnostic tools and can be used to make real time predictions.

The short time evolution used here is not only a requirement for
predicting chaotic systems but it also guarantees the consistency
of  our conclusions. This comes about since the primary step is to
identify  spatial points of small BV dimension and then to make
short time forecasts for the variables at these points. If the
dimension changed substantially during  the short time evolution
then the relationship between dimension and  prediction could not
be maintained. In this work predictions are made over  intervals
of 20  or 25 units of time (cycles) and under such circumstances
the value of BV dimension is practically constant.

For deterministic evolution some systems are more predictable than
others and this can be measured by Lyapunov exponents. However,
these exponents are well defined only asymptotically and are not
unique for finite time calculations \cite{abarbanel:91-01,
abarbanel:91-02, kurths:91-01, yoden:93-01}. In such cases bred
vectors are the proper tool to use and in  this paper we provided
the connection between predictability and the value of the BV
dimension.

The concept of bred vectors is intimately related to the analysis
of geophysical  flows \cite{patil:prl:2001, kalnay:book:2002}.
Other forecasting techniques could be envisaged for real
applications other than the CWM \cite{sauer:94-01:p, wan:94-01:p}.
The main thesis of this work however will remain  unaltered in
such situations since the system is deterministic and the simple
prediction algorithm or the  CWM are essentially finite
dimensional maps. Other algorithms would at most imply some
differences for instance in the error functions with no impact in
the main  conclusions. In a further investigation we intend to
apply the concept of BV dimension and the tools used herein to
several configurations of fluids in the turbulent regime.

\acknowledgments

This work is supported by Funda\c{c}\~ao de Amparo \`a  Pesquisa
do Estado de S\~ao Paulo (FAPESP), Brasil.


\end{document}